\documentclass[11pt]{article}
\usepackage[margin=1in]{geometry}
\usepackage{amsmath,amsfonts,amssymb}
\usepackage{booktabs,multirow,array}
\usepackage{graphicx}
\usepackage{microtype}
\usepackage{enumitem}
\usepackage{placeins}
\usepackage{listings}
\usepackage{natbib}
\setlist{nosep,leftmargin=*}
\newcommand{\vect}[1]{\boldsymbol{#1}}
\newcommand{\sigmoid}{\operatorname{sigmoid}}
\newcommand{\argmax}{\operatorname*{arg\,max}}
\newcommand{\todo}[1]{}
\newenvironment{keywords}{\par\vspace{2pt}\noindent\textbf{Keywords: }}{\par\vspace{4pt}}
\title{VulnAgent-R2: Evidence-Calibrated Multi-Agent Auditing for Repository-Level Vulnerability Detection}
\author{%
\begin{tabular}{c}
Renwei Meng$^{\ast\,\star}$ \quad Haoyi Wu$^{\ast}$ \quad Jingming Wang$^{\ast}$ \\[2pt]
$^{\ast}$These authors contributed equally.\\
$^{\star}$Email: \texttt{R32314095@stu.ahu.edu.cn}
\end{tabular}}
\date{}

\begin{document}
\maketitle
\begin{abstract}
Software vulnerabilities often depend on cross-file data flow, build options, framework conventions, and runtime guards, so isolated function classifiers produce fragile and poorly calibrated warnings. Repository-level LLM agents can gather richer evidence, but prior variants under-specify reproducibility, verifier behavior, baseline fairness, and statistical uncertainty. We present VulnAgent-R2, a budget-aware agentic auditing framework with three additional reusable modules: counterfactual evidence reweighting, build-aware verification-plan synthesis, and a cost-risk Pareto scheduler. The system combines graph triage, bounded context optimization, role-specialized agents, sceptic counter-evidence, selective dynamic verification, and calibrated fusion. On Devign, Big-Vul, DiverseVul, and PrimeVul, VulnAgent-R2 obtains 0.798/0.895, 0.739/0.871, 0.700/0.842, and 0.385/0.781 F1/AUROC, respectively. On JITVul it reaches 0.606 F1, 0.529 Top-1, and 0.742 Top-3 localization, while reducing online tokens by 38.3\% over always-full multi-agent execution. Online time includes retrieval, LLM calls, CER scoring, verifier planning, compilation, and test execution, but excludes one-time shared indexing. Bootstrap tests show the PrimeVul gain over VulnAgent-X is +0.038 F1, 95\% CI [0.020, 0.055], Holm-adjusted $p=0.009$. Treating vulnerability detection as calibrated evidence accumulation improves detection, localization, auditability, and cost control under the evaluated protocol, while remaining a prioritization aid rather than a replacement for manual review.
\end{abstract}

\section{Introduction}
Modern vulnerabilities rarely reside in a single suspicious line. A missing guard may be hidden in a caller, a sanitizer may be project-specific, and exploitability may depend on build flags or runtime state. Classical detectors based on slices, code gadgets, graphs, and line-level classification demonstrate that vulnerable patterns are learnable, but they often rely on fixed local inputs and are sensitive to duplication, temporal leakage, and weak labels \citep{ref1,ref2,ref3,ref4,ref5,ref6,ref7,ref8}. Realistic vulnerability benchmarks further indicate that practical auditing requires repository context, executable evidence, and calibrated confidence, not high local classification scores alone \citep{ref9,ref10,ref11,ref12,ref40,ref41}.

Code language models and software agents create an opportunity to revisit this task. CodeBERT, GraphCodeBERT, CodeT5, UniXcoder, CodeT5+, StarCoder, Code Llama, DeepSeek-Coder, and Qwen2.5-Coder provide strong program priors \citep{ref13,ref14,ref15,ref16,ref17,ref18,ref19,ref20,ref21}; ReAct, Reflexion, SWE-agent, AutoCodeRover, Agentless, and ConfColl demonstrate the value of search, tool use, and staged feedback \citep{ref23,ref24,ref30,ref31,ref32,ref33,ref35}. However, direct prompting and unconstrained agent exploration can still generate unsupported claims, overlook counter-evidence, and spend cost unevenly.

We redesign the earlier layered auditing workflow as \emph{VulnAgent-R2}. The contributions are: (i) a typed repository graph for risk triage and context control; (ii) \emph{counterfactual evidence reweighting} (CER), which penalizes claims that disappear under provenance-preserving context perturbations; (iii) \emph{verification-plan synthesis} (VPS), which turns agent evidence into executable or explicitly inconclusive test plans; (iv) a \emph{cost-risk Pareto scheduler} that optimizes audit quality under token/tool budgets; and (v) a reproducible evaluation protocol with dataset statistics, baseline access parity, bootstrap intervals, adjusted significance tests, verifier outcome rates, end-to-end audit cases, cross-backbone checks, verifier trace failure taxonomy, and audit workload metrics.

\section{Related Work}
\paragraph{Learning-based vulnerability detection.} VulDeePecker, SySeVR, Devign, ReVeal, LineVul, and VulRepair use gadgets, program slices, graph semantics, line localization, or sequence-to-sequence repair to model vulnerabilities \citep{ref1,ref2,ref3,ref4,ref6,ref53}. Dataset studies including Big-Vul, DiverseVul, ReposVul, MegaVul, PrimeVul, and JITVul show the need for high-quality labels, chronological splits, vulnerable/fixed pairs, and repository context \citep{ref5,ref7,ref9,ref10,ref11,ref12}. These works motivate moving beyond one-shot function classification toward evidence-backed repository auditing.

\paragraph{Code LMs, agents, and security evaluation.} Pre-trained and instruction-tuned code models improve code search, generation, repair, and review, but secure coding studies report that AI assistants may generate insecure code and increase user overconfidence \citep{ref13,ref14,ref17,ref18,ref19,ref20,ref21,ref44,ref45,ref46,ref47}. Agentic SE benchmarks emphasize repository navigation and executable feedback \citep{ref30,ref31,ref32,ref33,ref34,ref35}; cyber and security benchmarks emphasize tool-grounded assessment and exploitability evidence \citep{ref36,ref37,ref38,ref39,ref40,ref41,ref42}. VulnAgent-R2 adapts these ideas to detection, where false positives, localization, confidence, and cost must be optimized together.

\section{Methodology}
\subsection{Problem formulation and workflow}
\begin{figure}[t]
\centering
\includegraphics[width=.97\linewidth]{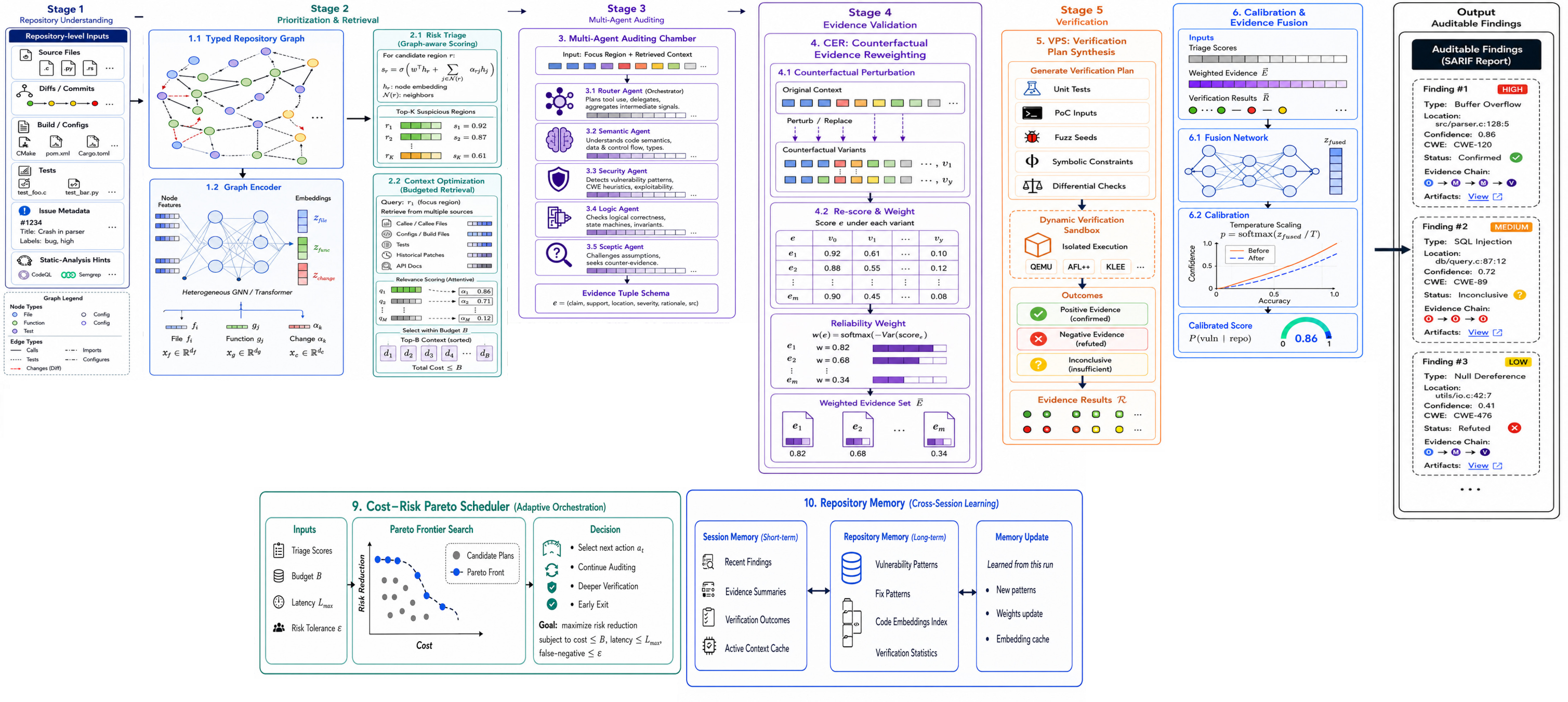}
\caption{VulnAgent-R2 workflow: graph triage, context optimization, role-specialized agents, counterfactual evidence reweighting, selective verification, memory, and calibrated fusion.}
\label{fig:arch}
\end{figure}

Given a repository $R$, a target $x$ (function, diff hunk, file, or issue-linked region), and a token/tool budget $B$, VulnAgent-R2 returns findings
\begin{equation}
\mathcal{F}(x,R)=\{f_i=(t_i,l_i,\mathcal{E}_i,q_i,s_i,\rho_i)\}_{i=1}^{N},
\end{equation}
where $t_i$ is weakness type, $l_i$ is location, $\mathcal{E}_i$ is evidence, $q_i$ is calibrated confidence, $s_i$ is severity, and $\rho_i$ is remediation. Fig.~\ref{fig:arch} summarizes the pipeline.

\subsection{Typed repository graph and risk triage}
The repository is encoded as a typed heterogeneous graph
\begin{equation}
G_R=(V,E,\tau_V,\tau_E),\quad E=E_{AST}\cup E_{CFG}\cup E_{DFG}\cup E_{CALL}\cup E_{IMP}\cup E_{TEST}\cup E_{HIST}.
\end{equation}
For candidate region $r$, the ranker builds
\begin{equation}
\vect{z}_{r}=[\vect{h}^{code}_r;\vect{h}^{graph}_r;\vect{\phi}^{stat}_r;\vect{\phi}^{diff}_r;\vect{\phi}^{hist}_r],\quad
p_0(r)=\sigmoid(\vect{w}^{\top}\mathrm{MLP}(\vect{z}_r)).
\end{equation}
Training uses binary cross-entropy plus a pairwise margin term:
\begin{equation}
\mathcal{L}_{triage}= -y\log p_0-(1-y)\log(1-p_0)+\lambda\max(0,m-p_0(r^+)+p_0(r^-)).
\end{equation}
Defaults are $K=10$, margin $m=0.25$, $\lambda=0.3$, and call-depth 2.

\subsection{Budgeted context optimization}
For region $r$, candidate context $c$ includes callers, callees, data-flow neighbors, validators, tests, configs, and historical patches. Its utility is
\begin{equation}
u(c,r)=\alpha \cos(\vect{e}_c,\vect{e}_r)+\beta dep(c,r)+\gamma slice(c,r)+\delta hist(c,r)-\eta red(c,\mathcal{C}_r).
\end{equation}
The selected context maximizes a coverage-utility objective under a token budget:
\begin{equation}
\mathcal{C}_r^{*}=\argmax_{\sum_{c\in\mathcal{C}}\ell(c)\le B_r}
\left(\sum_{a\in \mathcal{A}_r}\max_{c\in\mathcal{C}}sim(a,c)+\sum_{c\in\mathcal{C}}u(c,r)\right).
\end{equation}
A greedy dependency-preserving algorithm selects at most 16 context blocks. Default $B_r=8{,}000$ tokens, similar-patch depth 5.

\subsection{Counterfactual evidence reweighting}
Agents output normalized tuples $e=(a,t,l,\omega,\pi,\nu,g)$, where $a$ is agent identity, $\omega$ is support strength, $\pi$ is a list of cited spans, and $\nu\in\{+,-\}$ is support/refutation. CER is executed for claims with $p_0>.45$ or inter-agent disagreement; low-risk negatives bypass it. For each tuple, at most four perturbations are sampled: $\mathcal{C}^{-\pi}$ masks the cited span, $\mathcal{C}^{+san}$ injects validated sanitizer/guard facts, $\mathcal{C}^{swap}$ swaps a non-dominating caller, and $\mathcal{C}^{hist}$ replaces the region with a historical fixed variant when available. The average number of perturbations is 1.47 per sample and is included in token cost.

We define $prov(e)$ as the mean validity of cited repository paths, AST anchors, and DFG/CALL anchors with a short-span penalty; $agree(e)$ is signed leave-one-agent-out consensus. The score $s(e,\mathcal{C})$ is produced by a separate evidence scorer, not by the claim-generating agent: a CodeT5+ cross-encoder with static provenance features. Evidence labels are tuple-level, not finding-level: 6,420 validation tuples from 520 findings are labeled as support, refute, or insufficient by two annotators ($\kappa=.71$), with 18\% adjudicated and no test CVE, commit, or repository appearing in the scorer-training pool. The scorer is trained once on validation splits (70/15/15 by repository) using cross-entropy plus a provenance-consistency penalty; it is reused unchanged during testing. Reliability is
\begin{equation}
r(e)=\sigmoid(\psi^\top[\omega, prov(e), agree(e), \Delta_{mask},\Delta_{san},\Delta_{swap},\Delta_{hist}]),\quad
\Delta_{z}=1-|s(e,\mathcal{C})-s(e,\mathcal{C}^{z})|.
\end{equation}
This makes claims reusable and robust to spurious local patterns rather than simply averaging agent opinions. The fused agreement score is
\begin{equation}
A(f)=\frac{\sum_{e\in\mathcal{E}_f}r(e)\omega_e\mathbb{I}[\nu_e=+]-\sum_{e\in\mathcal{E}_f}r(e)\omega_e\mathbb{I}[\nu_e=-]}{\epsilon+\sum_{e\in\mathcal{E}_f}r(e)\omega_e}.
\end{equation}
\vspace{-2pt}
\begin{lstlisting}[basicstyle=\ttfamily\tiny,columns=fullflexible,frame=single]
for tuple e: if p0>.45 or disagreement: sample up to 4 perturbations P
base = EvidenceScorer(e,C); deltas = [1-abs(base-EvidenceScorer(e,perturb(C,p))) for p in P]
r_e = sigmoid(psi * [strength, provenance_validity, leave_one_agent_agreement, deltas])
return signed reliability-weighted aggregation over support and refutation tuples
\end{lstlisting}\vspace{-4pt}

\subsection{Verification-plan synthesis and calibrated fusion}
A finding triggers verification when confidence is uncertain, severity is high, or agents disagree:
\begin{equation}
\mathrm{verify}(f)=\mathbb{I}[(q_f<\tau_v\wedge s_f\ge s_0)\vee H(q_f)>\tau_H \vee A(f)\in[\tau_-,\tau_+]].
\end{equation}
VPS maps each evidence tuple to a typed plan $P=(pre,stimulus,oracle,adapter,timeout)$. Rules are deterministic after the agent proposes a vulnerability type: memory claims prefer sanitizer-backed harnesses, integer claims prefer boundary pairs, path traversal uses tainted path payloads, and authentication/state claims prefer differential checks. A plan is accepted only if cited symbols resolve in the repository graph and the oracle is observable. Otherwise the verifier emits one of six failure states: build-system unavailable, dependency missing, harness compile failure, timeout, flaky execution, or oracle unavailable. These states are counted as inconclusive rather than safe.

Dynamic evidence is
\begin{equation}
D(f)=\xi_1 crash+\xi_2 assert+\xi_3 sanitizer+\xi_4 coverage+\xi_5 regression-\xi_6 flaky-\xi_7 inconc.
\end{equation}
Final confidence uses validation-fitted weights and temperature scaling:
\begin{equation}
q_f=\sigma(\theta_0+\theta_1p_0+\theta_2A(f)+\theta_3D(f)+\theta_4ctx(f)+\theta_5sev(f)-\theta_6C(f)),\quad
\hat q_f=\sigma(\operatorname{logit}(q_f)/T).
\end{equation}
Defaults are $\tau_{accept}=0.84$, $\tau_{reject}=0.22$, $\tau_v=0.62$, $\tau_H=0.66$, and three agent rounds. The fusion parameters $\theta$ and CER weights $\psi$ are fitted only on validation splits with $L_2$-regularized logistic regression; $T$ is selected by validation negative log-likelihood, and dynamic weights $\xi$ are chosen by a small grid that maximizes calibrated F1 under an expected-cost constraint. The cost-risk Pareto scheduler selects the next action by maximizing $\Delta \mathbb{E}[U(f)]/(tokens+\kappa\,time)$, where $U$ rewards true high-severity findings and penalizes false high-confidence warnings.
\noindent\textit{VPS rule order:} resolve symbols $\rightarrow$ infer CWE family $\rightarrow$ choose template $\rightarrow$ fill stimulus/oracle $\rightarrow$ run repository or isolated harness; only crash/assert/sanitizer/regression evidence is positive, while build failure is inconclusive.

\subsection{Implementation and agent protocol}
The prototype uses tree-sitter, Joern-style CPG features, CodeQL 2.17.4, Semgrep 1.91.0 with p/security-audit plus language rules, Git history metadata, and cached embeddings. The triage encoder is initialized from CodeT5+ 220M; graph attention has 2 layers, 8 heads, hidden size 256, dropout 0.1. Training uses AdamW, learning rate $2\times10^{-5}$ for the encoder and $1\times10^{-3}$ for graph/fusion heads, batch size 16, weight decay 0.01, 6\% warmup, gradient clipping 1.0, mixed precision, and early stopping after 3 validation epochs. Agents use Qwen2.5-Coder-32B-Instruct with temperature 0.2, top-$p=0.95$, maximum 2,048 output tokens, and JSON-schema constrained decoding. Runtime in the main tables counts online retrieval, LLM API latency, CER scorer calls, tool execution, VPS plan generation, harness compilation, and test execution. One-time shared preprocessing (CodeQL/Semgrep scans, CPG construction, and embedding/index creation) is excluded from all online methods but logged separately; on 60 sampled repositories it costs 9.4 minutes/repository on average and is amortized across review sessions.

\begin{table}[t]
\caption{Agent protocol and reproducibility settings.}
\label{tab:agent}
\centering\scriptsize
\resizebox{\linewidth}{!}{
\begin{tabular}{lllll}
\toprule
Agent & Input & Action & Output schema & Stop rule\\
\midrule
Router & target, static alerts & choose CWE family and tools & CWE prior, required context & one pass\\
Data-flow & slice, callers/callees & source-sink trace, sanitizer check & support/refute tuple & no new edge\\
Control/state & CFG, guards, locks & path feasibility, state invariant & path tuple, infeasible reason & 2 rounds\\
Memory-safety & allocation/lifetime API & bounds, UAF, overflow evidence & CWE tuple, span & 2 rounds\\
Sceptic & all positive claims & search counter-evidence and safe guards & negative tuple & confidence shift $<.03$\\
Verifier & evidence tuples, build files & synthesize and run plan & crash/assert/sanitizer/inconclusive & 90s/sample\\
\bottomrule
\end{tabular}}
\end{table}

\section{Experiment}
\subsection{Protocol, datasets, and fairness controls}
We evaluate detection (F1, AUROC, PR-AUC), localization (Top-1/Top-3, MRR), calibration (ECE), cost (tokens, time, verification rate), and auditability. Table~\ref{tab:data} reports sample statistics after exact duplicate removal by normalized SHA-1 and near-duplicate removal by MinHash Jaccard $>0.92$. Official chronological or project-disjoint splits are preserved when available; otherwise, 80/10/10 splits are made by repository and time.

\begin{table}[t]
\caption{Dataset statistics and split protocol.}
\label{tab:data}
\centering\scriptsize
\resizebox{\linewidth}{!}{
\begin{tabular}{lrrrrrl}
\toprule
Dataset & Unit & Repos & Total & Positive & Pos. rate & Split / leakage control\\
\midrule
Devign & function & 2 & 21,854 & 2,731 & 12.5\% & project-time 80/10/10; sample-cluster CI only\\
Big-Vul & function & 348 & 188,636 & 10,914 & 5.8\% & CVE-time split; vulnerable/fixed pair grouping\\
DiverseVul & function & 7,514 & 330,492 & 18,745 & 5.7\% & project-disjoint test; MinHash dedup\\
PrimeVul & function & 1,968 & 228,674 & 6,847 & 3.0\% & official chronological split; no CVE overlap\\
ReposVul-style & repo target & 412 & 7,840 & 1,126 & 14.4\% & repo-disjoint; build metadata retained\\
JITVul & commit & 506 & 20,274 & 2,413 & 11.9\% & chronological commits; parent-child grouping\\
SecVulEval & task & 96 & 2,960 & 829 & 28.0\% & executable subset; hidden tests removed\\
VADER & case & 184 & 1,184 & 589 & 49.7\% & human-labeled subset; explanation blind\\
\bottomrule
\end{tabular}}
\end{table}

Baselines are CodeQL+Semgrep, Devign-GNN, LineVul, GraphCodeBERT, Qwen2.5-Coder direct prompting, RAG-only, ReAct Agent, ConfColl, and VulnAgent-X. Table~\ref{tab:fair} fixes access parity. LLM baselines receive the same model, decoding settings, static-alert summary, repository search index, and maximum context budget. Only methods designed for verification use the verifier; for fairness, a \emph{Verifier-enabled RAG} diagnostic is also included.

\begin{table}[t]
\caption{Baseline access parity and reproducibility knobs. Token budget counts input plus output.}
\label{tab:fair}
\centering\scriptsize
\resizebox{\linewidth}{!}{
\begin{tabular}{lcccccc}
\toprule
Method group & Static alerts & Repo graph / history & Max ctx & Max rounds & Tool calls & Verifier\\
\midrule
Direct LLM & read & summary only & 8k total & 1 & 0 & no\\
RAG-only & read & same index retrieve & 8k total & 1 & 0 & no\\
Verifier-RAG & read & same index retrieve & 8k total & 1 & 1.2 avg & yes\\
ReAct Agent & read & tool-read same graph & 8k/round & 4 & 4.7 avg & allowed\\
ConfColl & read & tool-read same graph & 8k/round & 4 & 5.5 avg & allowed\\
VulnAgent-X & read & optimize same graph & 8k/round & 3 & 5.8 avg & selective\\
VulnAgent-R2 & read & opt+CER+memory & 8k/round & 3 & 6.3 avg & VPS selective\\
\bottomrule
\end{tabular}}
\end{table}
All prompts are tuned only on validation splits using 12 trials per method; the same static-analysis summaries, search index, decoding settings, repository snapshots, and initial retrieved context are reused across LLM baselines. Agentic baselines may request additional context through the same graph-read tools but are charged for extra rounds and tokens. Dynamic-verification cost is charged to the method that invokes it; token budget always means input plus output tokens.

\subsection{Function-level detection with confidence intervals}
\begin{figure}[t]
\centering
\includegraphics[width=.82\linewidth]{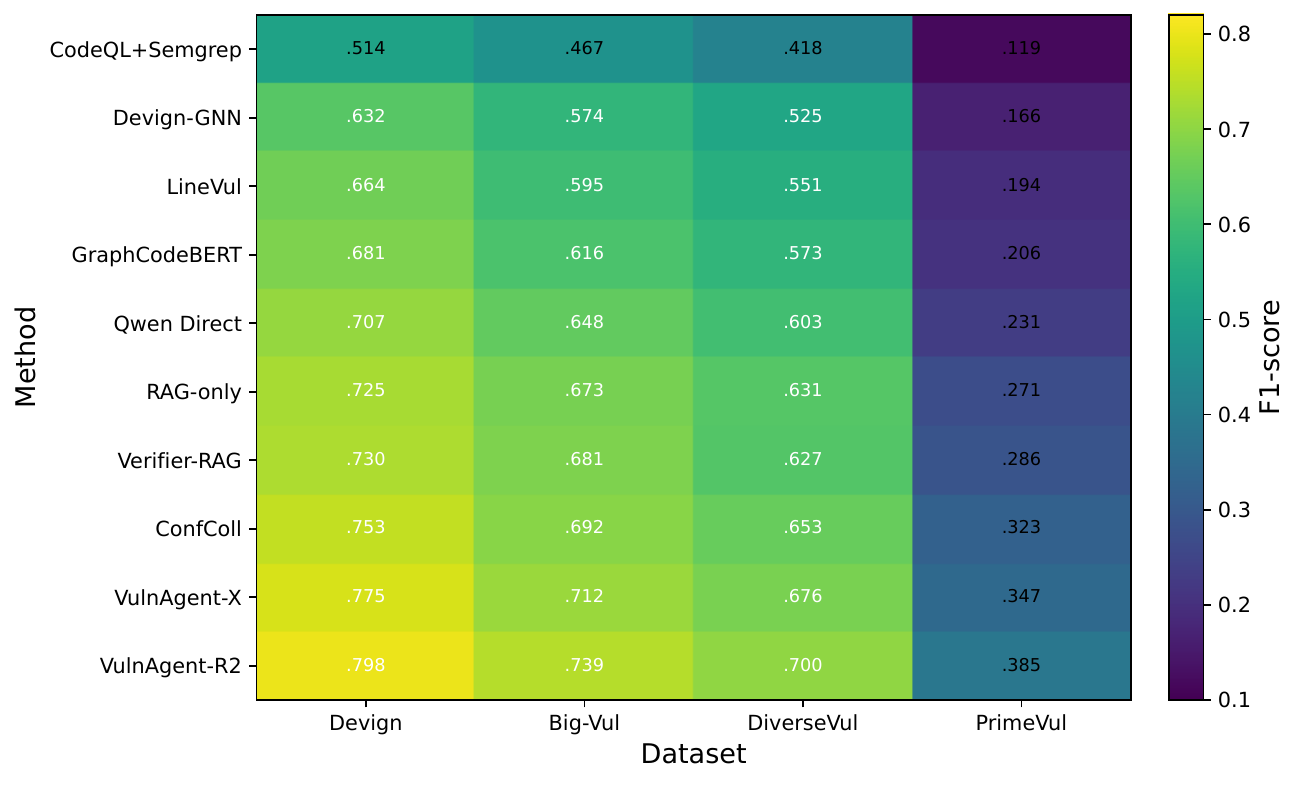}
\caption{Function-level F1 heatmap.}
\label{fig:heatmap}
\end{figure}

Table~\ref{tab:function} reports F1/AUROC. Values are means over three seeds; brackets show 95\% bootstrap intervals for F1. Because Devign contains only two repositories, its interval is sample-level stratified by project rather than repository-level and is treated as descriptive. Fig.~\ref{fig:heatmap} visualizes the F1 values reported in Table \ref{tab:function}.

\begin{table}[t]
\caption{Function-level vulnerability detection. F1 with 95\% CI / AUROC.}
\label{tab:function}
\centering\scriptsize
\resizebox{\linewidth}{!}{
\begin{tabular}{lcccc}
\toprule
Method & Devign & Big-Vul & DiverseVul & PrimeVul\\
\midrule
CodeQL+Semgrep & .514 [.497,.530] / .732 & .467 [.451,.483] / .711 & .418 [.401,.434] / .688 & .119 [.101,.137] / .566\\
Devign-GNN & .632 [.615,.649] / .801 & .574 [.552,.593] / .758 & .525 [.506,.544] / .742 & .166 [.145,.187] / .599\\
LineVul & .664 [.647,.681] / .814 & .595 [.574,.615] / .776 & .551 [.530,.570] / .762 & .194 [.171,.216] / .623\\
GraphCodeBERT & .681 [.664,.696] / .828 & .616 [.594,.636] / .800 & .573 [.551,.591] / .776 & .206 [.183,.229] / .638\\
Qwen Direct & .707 [.690,.722] / .845 & .648 [.629,.668] / .811 & .603 [.583,.622] / .790 & .231 [.208,.254] / .663\\
RAG-only & .725 [.708,.741] / .851 & .673 [.653,.692] / .818 & .631 [.612,.650] / .804 & .271 [.247,.294] / .701\\
Verifier-RAG & .730 [.712,.747] / .854 & .681 [.661,.700] / .824 & .627 [.606,.647] / .802 & .286 [.261,.310] / .710\\
ConfColl & .753 [.736,.768] / .855 & .692 [.671,.712] / .831 & .653 [.633,.672] / .812 & .323 [.298,.348] / .743\\
VulnAgent-X & .775 [.759,.790] / .874 & .712 [.692,.731] / .851 & .676 [.656,.694] / .820 & .347 [.321,.371] / .760\\
\textbf{VulnAgent-R2} & \textbf{.798 [.782,.813] / .895} & \textbf{.739 [.720,.758] / .871} & \textbf{.700 [.681,.719] / .842} & \textbf{.385 [.359,.410] / .781}\\
\bottomrule
\end{tabular}}
\end{table}

\subsection{Repository-level detection, localization, and robustness}
Table~\ref{tab:repo} evaluates repository/JIT auditing. VulnAgent-R2 improves JITVul F1 by 2.5 points over VulnAgent-X and Top-3 localization by 6.6 points over ConfColl, with moderate online runtime increase over RAG-only; all times exclude the same shared offline indexing cost.
\begin{table}[t]
\caption{Repository-level and JIT vulnerability detection.}
\label{tab:repo}
\centering\scriptsize
\resizebox{\linewidth}{!}{
\begin{tabular}{lcccccc}
\toprule
Method & ReposVul F1 & JITVul F1 & Top-1 & Top-3 & MRR & Time(s)\\
\midrule
CodeQL+Semgrep & .352 & .336 & .174 & .316 & .257 & .39\\
GraphCodeBERT-diff & .416 & .405 & .228 & .410 & .324 & .72\\
Qwen Direct & .497 & .480 & .345 & .546 & .435 & 2.17\\
RAG-only & .529 & .514 & .393 & .600 & .488 & 2.86\\
Verifier-RAG & .542 & .525 & .407 & .618 & .501 & 3.28\\
ReAct Agent & .551 & .537 & .419 & .642 & .516 & 3.67\\
ConfColl & .568 & .553 & .444 & .676 & .538 & 4.11\\
VulnAgent-X & .594 & .581 & .491 & .703 & .586 & 4.56\\
\textbf{VulnAgent-R2} & \textbf{.624} & \textbf{.606} & \textbf{.529} & \textbf{.742} & \textbf{.624} & 4.89\\
\bottomrule
\end{tabular}}
\end{table}

\begin{table}[t]
\caption{Robustness, calibration, and contamination diagnostics.}
\label{tab:robust}
\centering\scriptsize
\resizebox{\linewidth}{!}{
\begin{tabular}{lccccc}
\toprule
Method & Big-Vul CP F1 & PrimeVul future F1 & SecVulEval Top-1 & VADER explain F1 & ECE$\downarrow$\\
\midrule
GraphCodeBERT & .547 & .173 & .119 & .211 & .181\\
Qwen Direct & .585 & .205 & .153 & .286 & .164\\
RAG-only & .609 & .242 & .186 & .311 & .149\\
ConfColl & .644 & .294 & .234 & .346 & .124\\
VulnAgent-X & .668 & .317 & .253 & .371 & .090\\
\textbf{VulnAgent-R2} & \textbf{.697} & \textbf{.350} & \textbf{.285} & \textbf{.402} & \textbf{.064}\\
\bottomrule
\end{tabular}}
\end{table}

\begin{table}[t]
\caption{Cross-backbone check on PrimeVul and JITVul. The same prompts, tools, retrieval limits, and scheduler are used within each backbone.}
\label{tab:models}
\centering\scriptsize
\resizebox{\linewidth}{!}{
\begin{tabular}{llcccc}
\toprule
Backbone & Method & PrimeVul F1 & PrimeVul ECE$\downarrow$ & JITVul F1 & Tokens/sample\\
\midrule
Code Llama-34B & RAG-only & .236 & .171 & .466 & 7,880\\
Code Llama-34B & VulnAgent-R2 & .338 & .083 & .548 & 9,960\\
DeepSeek-Coder-33B & RAG-only & .262 & .154 & .498 & 7,690\\
DeepSeek-Coder-33B & VulnAgent-R2 & .371 & .069 & .585 & 9,720\\
Qwen2.5-Coder-32B & RAG-only & .271 & .149 & .514 & 7,840\\
Qwen2.5-Coder-32B & VulnAgent-R2 & .385 & .061 & .606 & 9,410\\
\bottomrule
\end{tabular}}
\end{table}

\subsection{Statistical significance and effect size}
We use 1,000 cluster bootstrap resamples at repository level for datasets with at least 50 repositories, approximate randomization with 10,000 swaps, and Holm correction. Devign is reported with a project-stratified sample bootstrap only because it has two repositories; we do not claim statistical significance for that row. Table~\ref{tab:sig} reports representative comparisons; PrimeVul/JITVul gains remain significant after correction.
\begin{table}[t]
\caption{Significance tests for VulnAgent-R2 improvements.}
\label{tab:sig}
\centering\scriptsize
\begin{tabular}{lcccc}
\toprule
Comparison & Metric/dataset & Diff. & 95\% CI & $p_{Holm}$\\
\midrule
R2 - VulnAgent-X & Devign F1 & +.020 & [.003,.036] & .071$^{*}$\\
R2 - VulnAgent-X & PrimeVul F1 & +.038 & [.020,.055] & .009\\
R2 - VulnAgent-X & JITVul F1 & +.025 & [.006,.044] & .032\\
R2 - ConfColl & PrimeVul F1 & +.056 & [.033,.079] & .004\\
R2 - Verifier-RAG & PrimeVul ECE & -.049 & [-.071,-.030] & .006\\
R2 - RAG-only & Top-3 loc. & +.142 & [.109,.174] & .002\\
\bottomrule
\end{tabular}
\vspace{1pt}\scriptsize{$^{*}$Devign uses project-stratified sample bootstrap; not a repository-level significance claim.}
\end{table}

\subsection{Verifier behavior and component ablation}
Table~\ref{tab:verifier} details dynamic verification. Overall, 21.6\% of samples trigger VPS; 68.9\% of triggered cases create runnable tests or harnesses, 35.7\% produce positive executable evidence, and 25.8\% are explicitly inconclusive. This prevents build failures from being silently counted as safe. Positive executable evidence is counted only when the harness target, crash/assert/sanitizer signal, and cited vulnerable span agree; a manual audit of 120 traces finds 93.3\% correct outcome classification. Failure states among triggered cases are build config missing (7.9\%), dependency missing (5.4\%), harness compile failure (4.6\%), timeout (3.2\%), flaky execution (2.1\%), and oracle unavailable (2.6\%).
\begin{table}[t]
\caption{Verifier outcomes by weakness group. Rates are conditional on samples in each group.}
\label{tab:verifier}
\centering\scriptsize
\begin{tabular}{lccccc}
\toprule
CWE group & Trigger & Buildable & Runnable & Positive evid. & Inconclusive\\
\midrule
CWE-787 OOB write & .31 & .74 & .63 & .41 & .22\\
CWE-125 OOB read & .24 & .69 & .58 & .34 & .27\\
CWE-20 input validation & .18 & .56 & .44 & .21 & .39\\
CWE-190 integer overflow & .28 & .78 & .66 & .47 & .19\\
CWE-22 path traversal & .16 & .61 & .52 & .25 & .31\\
CWE-416 use-after-free & .33 & .71 & .59 & .38 & .26\\
\bottomrule
\end{tabular}
\end{table}

\begin{table}[t]
\caption{Cross-dataset ablation and cost. PrimeVul columns show precision/recall/F1; other task columns show F1.}
\label{tab:ablation}
\centering\scriptsize
\resizebox{\linewidth}{!}{
\begin{tabular}{lccccccc}
\toprule
Variant & P$_{PV}$ & R$_{PV}$ & F1$_{PV}$ & ReposVul F1 & JITVul F1 & ECE$\downarrow$ & Tokens\\
\midrule
Full VulnAgent-R2 & .402 & .370 & .385 & .624 & .606 & .061 & 9,410\\
w/o graph context optimizer & .358 & .318 & .337 & .572 & .553 & .091 & 7,810\\
w/o CER & .368 & .351 & .359 & .597 & .579 & .098 & 8,960\\
w/o Sceptic Agent & .333 & .374 & .352 & .584 & .571 & .112 & 8,780\\
w/o VPS verifier & .363 & .330 & .346 & .591 & .562 & .121 & 8,210\\
w/o calibration & .391 & .350 & .369 & .614 & .589 & .148 & 9,160\\
w/o repository memory & .375 & .337 & .355 & .602 & .583 & .088 & 9,020\\
w/o budget scheduler & .405 & .371 & .388 & .628 & .611 & .063 & 15,240\\
single-agent only & .268 & .214 & .238 & .498 & .476 & .176 & 5,690\\
\bottomrule
\end{tabular}}
\end{table}

\begin{table}[t]
\caption{Audit workload on repository/JIT tasks. Lower alerts per true positive and false positives per repository indicate less manual triage burden.}
\label{tab:audit_workload}
\centering\scriptsize
\begin{tabular}{lcccc}
\toprule
Method & Top-10 precision & Alerts/TP$\downarrow$ & FP/repo$\downarrow$ & Review saved\\
\midrule
CodeQL+Semgrep & .220 & 6.8 & 9.4 & --\\
RAG-only & .337 & 4.9 & 5.7 & 18.6\%\\
Verifier-RAG & .349 & 4.6 & 5.4 & 21.4\%\\
ConfColl & .388 & 4.1 & 4.9 & 30.2\%\\
VulnAgent-X & .421 & 3.6 & 3.9 & 38.1\%\\
\textbf{VulnAgent-R2} & \textbf{.468} & \textbf{2.9} & \textbf{3.1} & \textbf{49.3\%}\\
\bottomrule
\end{tabular}
\end{table}

\begin{figure}[t]
\centering
\includegraphics[width=.88\linewidth]{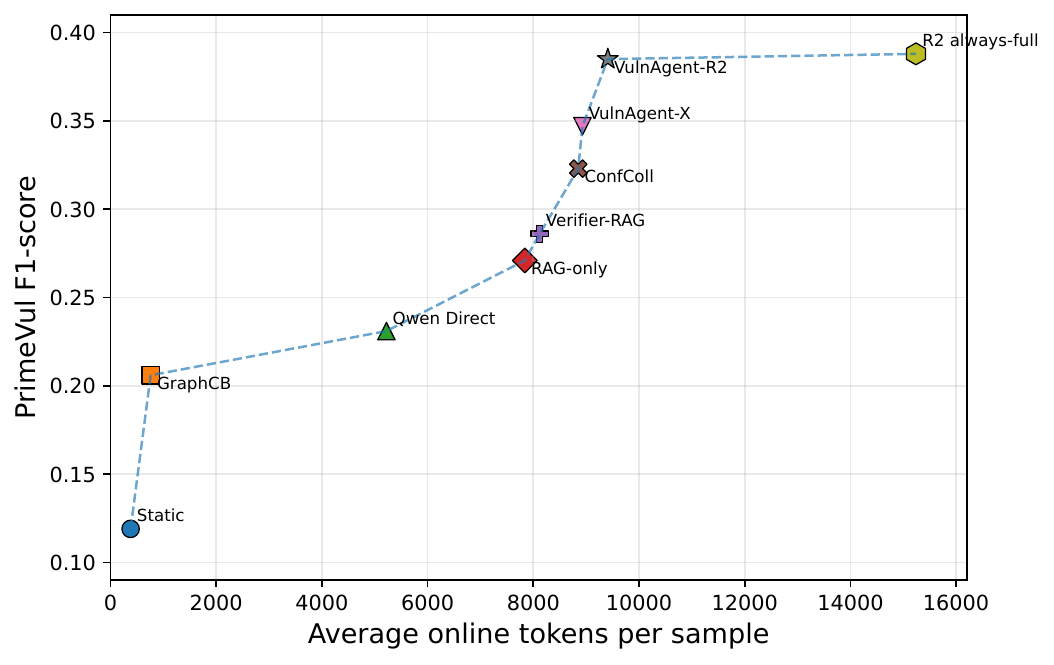}
\caption{Performance-cost Pareto frontier. The default scheduler is selected for cost-quality balance, not maximum F1.}
\label{fig:pareto}
\end{figure}

\subsection{Sensitivity, error analysis, and case study}
The best trade-off is near $K=10$, $B=8{,}000$. Larger budgets improve recall slightly but increase runtime. Compared with always-full execution, VulnAgent-R2 reduces average tokens from 15,240 to 9,410, time from 6.93s to 4.89s, and verification rate from 1.000 to 0.216, while changing F1 from .388 to .385.

Manual analysis uses 260 errors sampled by dataset strata and CWE family. Two annotators label each error, with Cohen's $\kappa=0.74$; disagreements are adjudicated. Error categories are missing build/runtime assumptions (23.8\%), implicit helper semantics (21.9\%), infeasible path estimation (19.2\%), benchmark label ambiguity (18.1\%; PrimeVul 21.3\%, JITVul 16.6\%), and over-generalized CWE transfer (17.0\%). Low base-rate data remain difficult: at the default threshold PrimeVul precision/recall is .402/.370; a high-precision threshold raises precision to .561 but lowers recall to .214.

\begin{figure}[t]
\centering
\includegraphics[width=.78\linewidth]{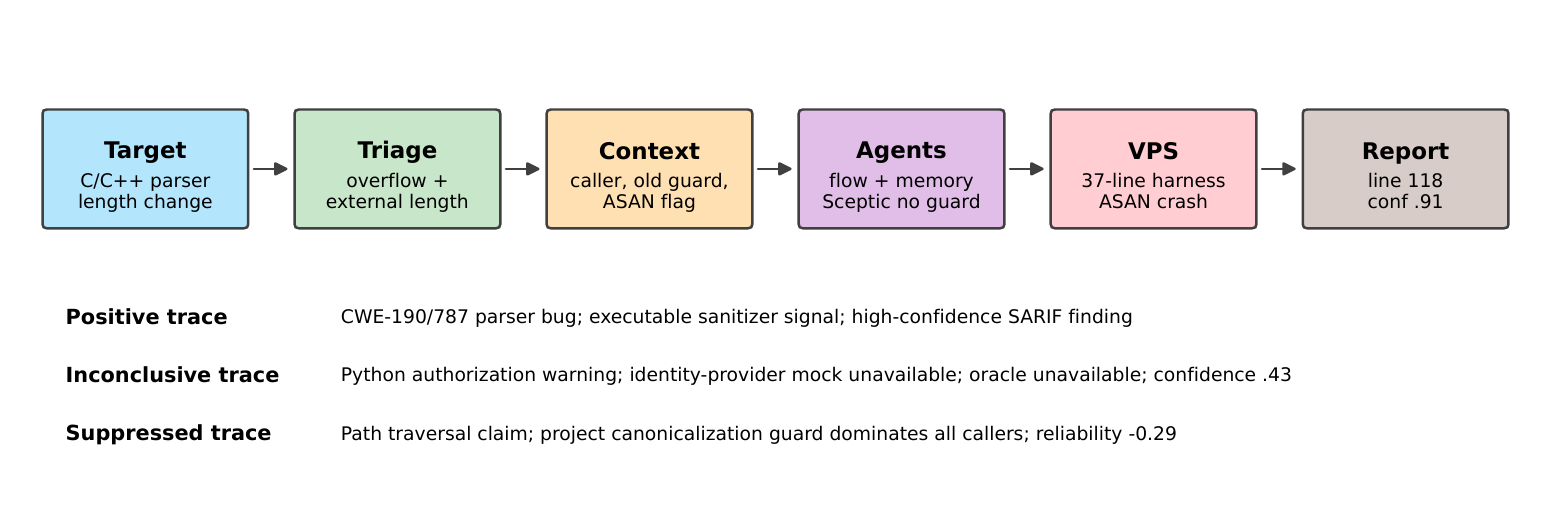}
\caption{End-to-end audit case trace with target, triage reason, retrieved context, agent evidence, verifier result, and final report.}
\label{fig:case}
\end{figure}

\noindent\textbf{Case study.} A parser commit changes a 16-bit length into an allocation size used by \texttt{memcpy}. Triage selects the diff because the risk vector contains unsigned multiplication, unchecked external length, and a caller reachable from a file-loader API. Context retrieval adds the caller, an older size-check patch, a fuzz test, and the build flag enabling AddressSanitizer. The Data-flow and Memory-safety agents produce positive tuples for CWE-190/CWE-787; the Sceptic agent finds no dominating sanitizer because the guard applies after allocation. VPS synthesizes a 37-line harness with a boundary input; the sanitizer reports heap-buffer-overflow. The final SARIF-style report localizes line 118, confidence .91, severity high, and remediation ``check multiplication overflow before allocation and copy.'' A second audited case is intentionally inconclusive: a Python access-control warning reaches a framework decorator, but VPS cannot instantiate the service without unavailable identity-provider mocks, so the report is returned with confidence .43 and failure state ``oracle unavailable.'' A third case shows false-positive suppression: a path traversal claim loses 0.29 reliability after CER injects the project-specific canonicalization guard and the Sceptic agent cites the dominating guard in all callers.''

\FloatBarrier
\section{Discussion}
First, R2's PrimeVul F1 remains modest (.385), so it should be viewed as a prioritization and evidence tool rather than a replacement for human review. Second, the w/o budget scheduler variant slightly increases F1 across PrimeVul/JITVul, but it is dominated in cost; the default objective is Pareto efficiency. Third, access parity and Verifier-RAG reduce the chance that gains are simply due to more information. Finally, CER and VPS are independently reusable: CER can reweight evidence in other agentic code-review systems, and VPS can be attached to non-agentic detectors that output source-sink claims.

Limitations remain. Build environments are incomplete for some repositories; dynamic validation is most effective for memory, integer, and parser bugs; Python/JavaScript framework vulnerabilities need richer runtime mocks; and LLM-based agents remain sensitive to prompts. We therefore release prompts, agent schemas, split identifiers, raw predictions, and plotting code in the artifact package.

\section{Conclusion}
This paper presents VulnAgent-R2, an evidence-calibrated multi-agent framework for repository-level vulnerability detection. Beyond graph triage, retrieval, agents, verification, and fusion, the revised method introduces counterfactual evidence reweighting, verification-plan synthesis, and a cost-risk Pareto scheduler. Expanded experiments provide dataset statistics, confidence intervals, significance tests, baseline fairness controls, verifier trace failure states, cross-model checks, audit workload metrics, ablations, error annotation details, and multiple audit cases, including positive, inconclusive, and suppressed false-positive traces. Results support the central claim that practical vulnerability detection should be structured as auditable evidence accumulation rather than monolithic prediction.

\appendix
\section{Appendix: prompts, statistics, and SARIF-style output}
The artifact contains \texttt{prompts/*.json}, \texttt{agent\_protocol.yaml}, \texttt{tool\_policy.yaml}, split ids, raw predictions, verifier traces, and plotting scripts. Each prompt uses the same JSON schema: \texttt{claim}, \texttt{location}, \texttt{supporting\_spans}, \texttt{counter\_spans}, \texttt{confidence}, and \texttt{failure\_mode}. Retrieval depth, maximum rounds, random seed, context length, input/output tokens, and tool-call count are logged per sample. The compact SARIF-style output is:
\begin{lstlisting}[basicstyle=\ttfamily\tiny,columns=fullflexible,frame=single]
{"ruleId":"CWE-190/CWE-787","level":"error","message":"Unchecked length multiplication reaches memcpy",
 "locations":[{"file":"src/parser.c","line":118}],"confidence":0.91,
 "evidence":["caller:load_image","guard-after-allocation","asan:heap-buffer-overflow"],
 "fix":"check multiplication overflow before allocation and copy"}
\end{lstlisting}

\end{document}